\documentclass[pdflatex,sn-mathphys-num]{sn-jnl}

\usepackage{graphicx}%
\usepackage{multirow}%
\usepackage{amsmath,amssymb,amsfonts}%
\usepackage{amsthm}%
\usepackage{mathrsfs}%
\usepackage[title]{appendix}%
\usepackage{xcolor}%
\usepackage{textcomp}%
\usepackage{manyfoot}%
\usepackage{booktabs}%
\usepackage{algorithm}%
\usepackage{algorithmicx}%
\usepackage{algpseudocode}%
\usepackage{listings}%
\usepackage{rotating}
\usepackage{tikz}
\usepackage{pgfplots}

\theoremstyle{thmstyleone}%
\theoremstyle{thmstyletwo}%

\theoremstyle{thmstylethree}%

\begin{document}

\title[An Initial Exploration of Fine-tuning Small Language Models for Smart Contract Reentrancy Vulnerability Detection]{An Initial Exploration of Fine-tuning Small Language Models for Smart Contract Reentrancy Vulnerability Detection}


\author*[1]{\fnm{Ignacio Mariano} \sur{Andreozzi Pofcher}}\email{ignacio.andreozzi.18@um.edu.mt}

\author[1,2]{\fnm{Joshua} \sur{Ellul}}\email{joshua.ellul@um.edu.mt}

\affil*[1]{\orgdiv{Centre for DLT}, \orgname{University of Malta}, \orgaddress{\city{Msida}, \country{Malta}}}

\affil[2]{\orgdiv{Department of Computer Science}, \orgname{University of Malta}, \orgaddress{\city{Msida}, \country{Malta}}}


\abstract{Large Language Models (LLMs) are being used more and more for various coding tasks, including to help coders identify bugs and are a promising avenue to support coders in various tasks including vulnerability detection --- particularly given the flexibility of such generative AI models and tools. Yet for many tasks it may not be suitable to use LLMs, for which it may be more suitable to use smaller language models that can fit and easily execute and train on a developer's computer. In this paper we explore and evaluate whether smaller language models can be fine-tuned to achieve reasonable results for a niche area: vulnerability detection --- specifically focusing on detecting the reentrancy bug in Solidity smart contracts.
}




\maketitle

\section{Introduction}\label{sec1}
Generative AI techniques have been proposed for various aspects of coding for tasks ranging from coding assistants \cite{millam2024coding} to optimisation \cite{krishna2024exploring} and vulnerability detection \cite{cheshkov2023evaluation} for which promising results are being heeded. Indeed, for many cases traditional types of code verification (be it at compile/development time \cite{ahrendt2019verification} or runtime \cite{ellul2018runtime}) often out perform generative AI-based techniques, yet such tools are often rigid and less flexible compared to how generative AI techniques can be used. Given potential future advancements of generative AI techniques, and given the flexible interface with which tools can interact with generative AI tools, it is useful to evaluate `how good are generative AI techniques at undertaking such tasks?' 

Indeed, extensive work in the domain has already been proposed surrounding this question, of which an extensive amount of literature has focused on the state-of-the-art large language models. Whilst it may be reasonable to make use of commercially/publicly available LLMs that are operated by a service provider, they indeed raise issues of privacy and confidentiality which some entities may rather not disclose certain intellectual property to (e.g. smart contract code). For this reason, we propose the use of small and resource-constrained language models for vulnerability detection that can execute on an individual's computer. In this paper we investigate the evaluation of small language models for a particular niche-area, i.e. for Solidity smart contracts specifically for the detection of a specific type of vulnerability --- the `reentrancy bug'.

The questions this paper aims to shed light on follow:
\begin{enumerate}
    \item Can small open-source language models (1-3B parameters) be fine-tuned to effectively detect reentrancy vulnerabilities in Solidity smart contracts?
    
    \item How do different model architectures (specifically LLaMA 3B and Qwen2.5Coder 3B) compare in their ability to adapt to the specialized task of reentrancy detection through parameter-efficient fine-tuning?
    
    \item Can synthetic data generation techniques produce training examples that enable effective model adaptation despite the scarcity of real-world vulnerability examples?
\end{enumerate}

The remainder of this paper is structured as follows. In Section~\ref{sec:dataset} we describe the curated dataset, and then in Section~\ref{sec:models} delve into details pertaining to the small language models. We then provide evaluation details of the fine-tuned small language models in Section~\ref{sec:eval}, and for provide a comparison relative to state-of-the-art large language models in Section~\ref{sec:llmeval}. We then conclude in Section~\ref{sec:conc}.

\section{Dataset Composition}\label{sec:dataset}
In this section, we detail the methodology adopted for generating the training and test datasets, highlighting the reasoning underpinning the selected class distributions. 

To develop an effective vulnerability detection model, comprehensive training and test datasets were constructed, considering class distributions. The intrinsic scarcity of vulnerable contracts within production environments naturally leads to highly imbalanced datasets when gathering real-world samples. This imbalance can result in models achieving deceptively high accuracy simply by always predicting the predominant (secure) class. Custom-crafted balanced datasets mitigate this issue by guaranteeing an equal representation across vulnerability classes during the model's training phase.

The training dataset developed specifically for this study consisted of 8,000 Solidity smart contracts, carefully balanced between 4,000 contracts exhibiting reentrancy vulnerabilities and 4,000 secure contracts without such vulnerabilities. Of the vulnerable contracts, 7.5\% (300) were sourced from the Reentrancy Study Dataset \cite{zheng2023turn} and manually modernised through a process described in subsequent sections, whilst the remaining 92.5\% (3,700) were systematically synthesised through controlled generation methods. Likewise, within the secure subset, 10\% (400) originated from verified secure examples within the Reentrancy Study Dataset, and the remaining 90\% (3,600) were synthesised using template-based approaches implementing various security patterns. The predominantly synthetic nature of the dataset was necessitated by the notable scarcity of well-documented, sophisticated and contemporary instances of reentrancy vulnerabilities available in public repositories.

The test dataset—used initially to evaluate baseline model performance and subsequently to benchmark the trained model—comprised 120 Solidity smart contracts, constructed using a stratified sampling procedure to ensure representative coverage of both vulnerability-free and vulnerability-containing instances. This holdout dataset, representing approximately 1.5\% of the overall corpus, preserved a near-balanced class distribution, with 47.5\% (57 contracts) containing reentrancy vulnerabilities and 52.5\% (63 contracts) deemed secure. This distribution was a deliberate design decision aimed at reducing evaluation bias while retaining alignment with real-world vulnerability prevalence trends.

The composition of the test set follows a hybrid approach to evaluation data curation. Of the 57 contracts (47.5\%) containing reentrancy vulnerabilities, 44 were sourced from the Reentrancy Study Dataset \cite{zheng2023turn}, while the remaining 13 represented documented exploits observed in production environments, drawn from Caversaccio's curated repository\footnote{\url{https://github.com/pcaversaccio/reentrancy-attacks}}. 

To address limitations of limited labeled data, outdated solidity versions and issues emanating due to language changes, this research adopted a multi-faceted approach to data collection and refinement, integrating source code repository extraction, expert annotation, and systematic preprocessing. The 4,000 vulnerable contracts used for training were curated through a stratified sampling methodology: 300 contracts (7.5\%) were sourced from the Reentrancy Study Dataset (as discussed) while the remaining 92.5\% were synthetically generated by implementing parameterised vulnerability patterns. Notably, the 300 contracts drawn from the Reentrancy Study Dataset required substantial modification prior to inclusion in the training set due to their reliance on legacy Solidity versions (primarily 0.4.x and 0.5.x) that made use of deprecated external call mechanisms such as \texttt{transfer()} and \texttt{send()}, and lacked explicit overflow protection features introduced in version 0.8.0. 

The updating process involved manually adapting the contracts to align with modern Solidity standards (version 0.8.0 and above), ensuring that the original vulnerabilities were preserved while rendering the code reflective of contemporary development practices. This transformation included replacing deprecated constructs (e.g., substituting \texttt{transfer()} with \texttt{call\{value: ...\}("")}), introducing explicit variable visibility modifiers, and revising arithmetic operations to account for the integrated overflow checks present in newer Solidity versions. The modernisation effort was carried out through a combination of manual contract-level review and programmatic transformation techniques.

This modernisation process also involved incorporating explicit visibility modifiers for variables and functions, as well as adapting arithmetic operations to leverage the built-in overflow protection introduced in Solidity 0.8.0—thereby ensuring consistency with modern security conventions. The same modernisation protocol was applied to the 400 secure contracts sourced from the Reentrancy Study Dataset, bringing the total number of incorporated contracts from this dataset to 700 (300 vulnerable and 400 secure), collectively contributing approximately 8.75\% of the overall training corpus. While quantitatively modest, this subset played a critical role in anchoring the dataset in empirically validated vulnerability instances and informing the parameterisation of synthetically generated samples. The comprehensiveness of the Reentrancy Study Dataset—achieved through a hybrid labeling methodology combining static analysis, dynamic execution, and expert verification—provided a robust foundation for the development of reliable vulnerability detection models. Temporal disparities, however, posed a challenge, the smart contracts were collected between 2015 and 2022, with over 63\% predating Solidity 0.8.0. As a result, a systematic modernisation method was followed to ensure consistency with contemporary language standards.

\subsection{Synthetic Data generation}
We now present the methodology employed for generating synthetic smart contract data, with the aim of producing diverse, representative, and structurally valid samples to support robust model training.

Given the limited availability of real-world examples of reentrancy vulnerabilities—with only 147 documented exploits identified in the comprehensive repository curated by Caversaccio\footnote{\url{https://github.com/pcaversaccio/reentrancy-attacks}}, and merely 13 exhibiting sufficiently isolated vulnerability patterns suitable for training --- synthetic data generation became a foundational pillar of the dataset construction strategy. This approach is consistent with methodologies advocated by Godefroid et al.~\cite{godefroid2017learn} and Hellendoorn et al.~\cite{hellendoorn2019global}, who proposed synthetic generation as a viable means to mitigate data scarcity in program analysis domains. Even the extraction and preparation of this limited subset demanded considerable effort, as vulnerability-containing contracts often required extensive disentanglement from surrounding contract ecosystems to isolate the vulnerable components.

The dataset size of 8,000 contracts (4,000 vulnerable and 4,000 non-vulnerable) was determined based on empirical evidence concerning the relationship between dataset scale and model performance in specialised classification tasks. 

The synthetic data generation methodology employed multiple techniques to ensure both diversity and representativeness. We adopted a template-based generation strategy with controlled parameterisation, maintaining consistency in fundamental vulnerability patterns while introducing substantial variation in surface-level features. To address class imbalance characteristics of real-world vulnerability distributions, we employed strategic oversampling techniques, including SMOTE (Synthetic Minority Over-sampling Technique), originally proposed by Chawla et al \cite{chawla2002smote}. 

For smaller language models (1-3B parameters), a dataset size of 8,000 examples constitutes an appropriate scaling factor.

\subsection{Pattern based Generation}
The first generation technique yielded 2,800 vulnerable contracts exhibiting basic reentrancy patterns, generated through controlled parameterisation of fundamental vulnerability templates. This method focused on producing variants of elementary reentrancy vulnerabilities by systematically randomising variable names, function structures, and control flow constructs, while preserving the underlying vulnerability semantics. The implementation incorporated semantic-preserving transformations, ensuring that the generated contracts retained essential vulnerability characteristics while introducing surface-level diversity necessary to mitigate overfitting to superficial code patterns.

\subsection{Advanced Vulnerable Contracts}
The second generation technique produced an additional 900 vulnerable contracts, each implementing more sophisticated vulnerability patterns through a taxonomy-guided generative framework. This methodology adopted a systematic approach to generating contracts across four distinct reentrancy vulnerability types: single-function reentrancy, cross-function reentrancy, cross-contract reentrancy, and read-only reentrancy. The implementation incorporated randomised naming for functions and contracts to prevent the model from learning spurious textual cues, while preserving the structural features that define each vulnerability subtype.

To address potential class imbalance among more complex reentrancy variants, this technique leveraged SMOTE, to ensure an even distribution across vulnerability subtypes. An illustrative example of the parameterised template-based generation process for the single-function reentrancy subtype is provided in Figure~\ref{code:single_reentrancy}.

\begin{figure}[H]  
\caption{Single-Function Reentrancy Vulnerability Template}
\label{code:single_reentrancy}
\begin{verbatim}
function generate_solidity_contract(vuln_type):
contract_name = "VulnContract" + str(random.randint(1000, 9999))
function_name = generate_random_function_name()

if vuln_type == "single_function_reentrancy":
    contract_code = f"""
    pragma solidity ^0.8.0;

    contract {contract_name} {{
        mapping(address => uint256) public balances;

        function deposit() public payable {{
            balances[msg.sender] += msg.value;
        }}

        function {function_name}() public {{
            require(balances[msg.sender] > 0, "Insufficient balance");
            (bool success,) = msg.sender.call{{value: balances[msg.sender]}}("");
            require(success, "Transfer failed");
            balances[msg.sender] = 0;
        }}
    }}
    """
\end{verbatim}
\end{figure}

As illustrated in Figure~\ref{code:single_reentrancy}, the template captures the core reentrancy vulnerability pattern --- the execution of an external call prior to the corresponding state update, which enables the reentrant behaviour. The parameterised components, such as the contract and function names, introduce surface-level variability while preserving the semantic structure of the vulnerability.

\subsection{Vulnerability-free Contracts}
For the generation of vulnerability-free contracts, two complementary techniques were employed. The first vulnerability-free contract generation technique yielded 2,800 contracts that implemented various security patterns specifically designed to mitigate reentrancy vulnerabilities. This methodology utilised multiple templates incorporating best practices such as the Checks-Effects-Interactions pattern, ReentrancyGuard implementations\footnote{\url{https://docs.openzeppelin.com/contracts/4.x/api/security\#ReentrancyGuard}}, pull-payment mechanisms and mutex locks.

Figure~\ref{code:ReentrancyGuard} illustrates one of the template categories employed specifically demonstrating the implementation of the \texttt{ReentrancyGuard} pattern as defined in the OpenZeppelin library.

\begin{figure}[H]
\caption{Reentrancy Guard Secure Contract Example}
\label{code:ReentrancyGuard}
\begin{verbatim}
contract_templates = [
"""
pragma solidity ^0.8.19;
import "@openzeppelin/contracts/security/ReentrancyGuard.sol";

contract SecureFund{0} is ReentrancyGuard {{
    mapping(address => uint256) private balances;

    function deposit() external payable {{
        require(msg.value > 0, "Must send ETH");
        balances[msg.sender] += msg.value;
    }}

    function withdraw(uint256 _amount) external nonReentrant {{
        require(balances[msg.sender] >= _amount, "Insufficient balance");
        balances[msg.sender] -= _amount;
        payable(msg.sender).transfer(_amount);
    }}
}}
"""
\end{verbatim}
\end{figure}

In Figure~\ref{code:ReentrancyGuard}, the critical security element is the \texttt{nonReentrant} modifier from OpenZeppelin's \texttt{ReentrancyGuard}, which enforces a mutex mechanism to prevent reentrant calls. This pattern exemplifies one of four security strategies systematically incorporated into the generated contracts.

\subsection{Advanced Secure Contracts}

The second secure contract generation technique yielded an additional 800 contracts with more sophisticated security implementations. This methodology focused on constructing contracts exhibiting ``deceptive complexity'' --- i.e. contracts which may appear superficially vulnerable to static analysis tools but internally incorporate layered security mechanisms to defend against reentrancy attacks. These contracts implemented multiple security modifiers, including custom non-reentrancy locks, block execution limits, gas-based execution guards, timestamp throttling, and secure delegation checks.

Diversity in security implementation techniques was essential for training the model to recognise a broad spectrum of secure coding patterns, rather than relying on simplistic indicators of vulnerability absence. Without exposure to varied and realistic security architectures, the model risks developing oversimplified heuristics for vulnerability detection.

\subsection{Integration of Real-World Vulnerabilities}
The integration of real-world vulnerability instances into the testing dataset is critical for ensuring a realistic evaluation of the model's detection capabilities. In contrast to purely synthetic datasets --- which may overlook the nuanced characteristics of practical exploits, real-world vulnerabilities serve as authentic, adversarial examples that more accurately reflect deployment conditions and challenge the model's robustness in realistic scenarios.

Building upon the method outlined above, the testing dataset used for the evaluation of both baseline and trained models was strategically enhanced through the deliberate inclusion of empirically documented real-world reentrancy vulnerabilities. Of the 120 smart contracts comprising our testing dataset, 13 were directly derived from empirically verified reentrancy exploits observed in production blockchain environments, several of which were recorded as recently as late 2024.

The inclusion of recent exploitation incidents such as the Peapods Finance attack, The Smoofs attack and the Sumer Money attack, ensures that our testing framework evaluates model performance against contemporary attack methodologies that have demonstrably bypassed existing security protocols.

\subsection{Reentrancy Variants}
Our dataset design ensures comprehensive evaluation across reentrancy variants including instances of five principal reentrancy types: Single-Function Reentrancy, Cross-Function Reentrancy, Cross-Contract Reentrancy and Read-Only Reentrancy.

\subsection{Test Dataset Construction}
The construction of the test dataset adhered to a meticulous three-phase qualification process designed to ensure accuracy, diversity and compatibility (with 0.8.0 solidity versions).

\paragraph{1. Static Analysis Verification}
Candidate contracts underwent analysis using the static analysis tool \texttt{Slither} to establish their classification as either vulnerable or secure. To ensure confidence in the results contracts with conflicting analysis outcomes were either excluded or subjected to further manual review to resolve ambiguities.

\paragraph{2. Syntactic Modernisation}
Contracts originating from the Reentrancy Study Dataset were modernised to ensure compatibility with Solidity 0.8.0. Key updates included updating pragma directives, replacing deprecated functions, adding explicit visibility modifiers and refactoring arithmetic operations to align with Solidity's integrated SafeMath features.

\paragraph{3. Structural Diversity Assurance}
The test set was carefully curated to encompass different reentrancy vulnerabilities, including single-function reentrancy, cross-function reentrancy, cross-contract reentrancy and read-only reentrancy.

\paragraph{4. Additional Checks}
All selected contracts underwent cross-validation using OpenAI language models. This step leveraged advanced contextual understanding to mitigate potential oversights from earlier phases.

\paragraph{5. Manual Review and Checks}
Finally, manual review to double-check accurate classification as vulnerable or non-vulnerable was undertaken. 

\subsection{Modernization of Reentrancy Study Dataset Contracts}
Although the Reentrancy Study Dataset formed the foundation of our test set, as discussed, substantial improvements were made to enhance its relevance and robustness including replacing deprecated Solidity functions, implementing explicit overflow/underflow protection mechanisms, and refactoring control flow structures to comply with Solidity 0.8.x standards.

To address limitations in prior datasets, we implemented several key improvements including synthesized contracts were added, and we increased the sample size of read-only reentrancy samples.

\section{Model Selection \& Fine-Tuning Strategy of Models: LLaMA 3B, Qwen2.5Coder 3B}
In this section we discuss why we decided to evaluate in this paper LLaMA 3B and Qwen2.5Coder 3B as the smaller models for reentrancy detection in this first study.

\subsection{LLaMA 3.2 3B Model Architecture}\label{sec:models}
The LLaMA 3.2 3B model represents a significant advancement in the deployment of transformative AI capabilities for relatively resource-constrained environments. This model serves as a compact yet capable member of the LLaMA family, engineered specifically for scenarios with limited computational resources while maintaining strong language understanding performance\footnote{\url{https://huggingface.co/stromdotcom/Llama-3.2-3B-Instruct-tuned}}.

LLaMA 3.2 3B utilises a decoder-only transformer architecture comprising approximately 3 billion parameters. The model integrates advanced attention mechanisms and has undergone extensive pre-training on diverse corpora, including both general natural language and code. Building upon the foundational advances of earlier LLaMA iterations, it incorporates architectural refinements that improve its capacity to process and reason about structured text, such as programming languages.

We selected the LLaMA 3.2 3B model as one of the models to evaluate based on key factors:
\begin{itemize}
    \item \textbf{Capability vs. Efficiency:} The 3B scale balances complex pattern understanding with deployability on consumer hardware.
    \item \textbf{Instruction-following:} Pre-training on instruction datasets enables strong zero-shot performance in specialized tasks.
    \item \textbf{Quantization Suitability:} The architecture maintains performance under quantization, significantly reducing memory demands.
\end{itemize}

\subsection{Qwen2.5-Coder-3B Architecture}
Qwen2.5-Coder is a specialised foundation model optimised for code understanding and generation, rendering it particularly well-suited for security-related analysis tasks and it has underwent extensive pre-training on 5.5 trillion tokens, with a substantial portion of the corpus dedicated to diverse programming languages~\cite{hui2024qwen2}.

Qwen2.5-Coder-3B integrates key enhancements over general-purpose models and was selected as another model to investigate based on the following:
\begin{itemize}
    \item \textbf{Code-specific attention:} Optimised attention mechanisms tailored to the syntactic and semantic structure of programming languages.
    \item \textbf{Enhanced tokenization:} Utilises a tokeniser designed to preserve meaningful code constructs, improving parsing and comprehension.
    \item \textbf{Instruction-following:} Fine-tuned to follow complex code analysis directives, enabling effective handling of specialised security tasks.
\end{itemize}

\subsection{Quantization Implementation}
The implementation employs Unsloth's dynamic 4-bit quantization, representing a significant advancement over traditional quantization techniques. Whereas conventional 4-bit quantization frequently results in unacceptable accuracy degradation, Unsloth's approach mitigates this by selectively excluding parameters from quantization based on their sensitivity to precision loss\footnote{\url{https://unsloth.ai/blog/dynamic-4bit}}.

\subsection{Parameter-Efficient Fine-Tuning with LoRA}
Low-Rank Adaptation (LoRA) is employed as the primary fine-tuning method, enabling parameter-efficient adaptation by introducing trainable low-rank matrices into the transformer architecture while keeping the pre-trained weights frozen\footnote{\url{https://research.ibm.com/blog/LoRAs-explained}}. This approach significantly reduces memory consumption, as gradients and optimizer states are computed solely for the LoRA parameters.

The fine-tuning process focuses on adapting the attention mechanisms and output projection layers for the vulnerability detection task, leveraging the pretrained knowledge embedded within the model. This targeted adaptation enables the model to associate semantic code patterns with security implications, without requiring extensive retraining. 

To further optimise memory usage, Unsloth's gradient checkpointing is applied, which reduces memory overhead by recomputing intermediate activations during backpropagation, trading-off increased computational complexity for reduced memory consumption. 

\section{Evaluation and Validation}\label{sec:eval}
In this section we evaluate the fine-tuned language models discussed in Section~\ref{sec:models}. 

The evaluation employed a comprehensive set of performance metrics to assess model effectiveness across multiple dimensions including accuracy, precision, recall, F1-scores and confusion matrices. The various metrics were computed with scikit-learn \cite{pedregosa2011scikit}. 

\subsection{Base Model Performance}
Before the parameter-efficient fine-tuning regimen described in Section~\ref{sec:models} was evaluated, the unaltered foundational models were tested and yielded poor performance as provided in Table \ref{tab:base_performance}. This baseline evidences the importance of domain-specific adaptation for such niche areas. 

\begin{table}[h]
\centering
\caption{Base Model Performance Metrics}
\label{tab:base_performance}
\begin{tabular}{lcccc}
\hline
\textbf{Model} & \textbf{Accuracy} & \textbf{Precision} & \textbf{Recall} & \textbf{F1-Score} \\
\hline
LLaMA 3B & 48\% & 0.46 & 0.47 & 0.46 \\
Qwen2.5Coder 3B & 45\% & 0.43 & 0.44 & 0.43 \\
\hline
\end{tabular}
\end{table}

The baseline results corroborate observations made by Rabin et al. \cite{rabin2021generalizability}, who documented substantial performance shortfalls when general-purpose code language models are deployed for specialised analysis without targeted adaptation. Notably, the Qwen 2.5 Coder 3B model exhibited marginally inferior baseline performance to the more general LLaMA 3B model, indicating that broad code-comprehension aptitude does not necessarily confer proficiency in niche areas like solidity vulnerability detection. 

\subsection{Fine-tuned Model Performance}

We now delve into performance achieved for the fine-tuned models in Sections~\ref{sec:evalllama} and \ref{sec:evalqwen}.

\subsubsection{Evaluating the Fine-tuned LLaMA 3B Model} \label{sec:evalllama}
By employing the LoRA-based parameter-efficient fine-tuning protocol, together with the synthetic-data augmentation workflow detailed in the above sections, the LLaMA 3B model's performance increased to 67\% test accuracy --- a 19-percentage-point gain over the baseline. The result is particularly notable given the constrained computational requirements imposed, indicating effective transfer of pretrained knowledge into the specialised, niche domain.

\begin{table}
  \centering
  \caption{Confusion matrix for fine-tuned LLaMA-3 (3 B) model}
  \label{tab:llama_confusion}

  \begin{tabular}{|c|c|c|}
    \hline
          & \multicolumn{2}{c|}{\textbf{Predicted}} \\ 
    \textbf{Actual} & \textbf{Non-Vulnerable} & \textbf{Vulnerable} \\ \hline
    \textbf{Non-Vulnerable} & 47 & 4  \\ \hline
    \textbf{Vulnerable}     & 26 & 15 \\ \hline
  \end{tabular}
\end{table}

Granular performance indicators are provided in Table \ref{tab:llama_metrics}.

\begin{table}[h]
\centering
\caption{Fine-tuned LLaMA 3B Performance Metrics}
\label{tab:llama_metrics}
\begin{tabular}{l c c c}
\toprule
\textbf{Class} & \textbf{Precision} & \textbf{Recall} & \textbf{F1-Score} \\
\midrule
Non-Vulnerable & 0.64 & 0.92 & 0.76 \\
Vulnerable & 0.79 & 0.37 & 0.50 \\
\midrule
Weighted Average & 0.71 & 0.67 & 0.64 \\
\bottomrule
\end{tabular}
\end{table}

The model's achieves a reasonable precision (0.79) for vulnerable class predictions --- which is a valuable characteristic for security analysis where false positives can be costly. Results acheived are comparable with many traditional approaches to vulnerability detection, though indeed still not as good as some traditional methods. The model exhibits ambiguity in 28 of the 120 contracts --- a phenomenon recognised in language-model applications to intricate technical tasks.

\subsubsection{Evaluating the Fine-tuned Qwen2.5Coder 3B Model} \label{sec:evalqwen}

The fine-tuned Qwen 2.5 Coder 3B model attained 59\% accuracy --- an uplift of 14 percentage points relative to the baseline, yet it still lagged behind the fine-tuned LLaMA 3B model.

\begin{table}[h]
  \centering
  \caption{Confusion matrix for fine-tuned Qwen2.5Coder 3B Model}
  \label{tab:qwen_confusion}

  \begin{tabular}{|c|c|c|}
    \hline
          & \multicolumn{2}{c|}{\textbf{Predicted}} \\ 
    \textbf{Actual} & \textbf{Non-Vulnerable} & \textbf{Vulnerable} \\ \hline
    \textbf{Non-Vulnerable} & 32 & 31  \\ \hline
    \textbf{Vulnerable}     & 18 & 39 \\ \hline
  \end{tabular}
\end{table}

The confusion matrix provided in Figure \ref{fig:qwen_confusion} for the Qwen2.5Coder 3B fine-tuned model reveals a less accurate classification pattern than the LLaMA 3B fine-tuned model, with 51\% of non-vulnerable contracts and 68\% of vulnerable contracts correctly identified. 

Inspection of the confusion matrix in Figure \ref{fig:qwen_confusion} shows that the fine-tuned Qwen 2.5 Coder-3 B offers a comparatively less accurate classification profile than the fine-tuned LLaMA-3B, correctly identifying 51\% of non-vulnerable contracts and 68\% of vulnerable ones.

\begin{table}[h]
\centering
\caption{Fine-tuned Qwen2.5Coder 3B Performance Metrics}
\label{tab:qwen_metrics}
\begin{tabular}{lccc}
\hline
\textbf{Class} & \textbf{Precision} & \textbf{Recall} & \textbf{F1-Score} \\
\hline
Non-Vulnerable & 0.64 & 0.51 & 0.57 \\
Vulnerable & 0.56 & 0.68 & 0.61 \\
\hline
Weighted Average & 0.60 & 0.59 & 0.59 \\
\hline
\end{tabular}
\end{table}

Table \ref{tab:qwen_metrics} indicates that the fine-tuned Qwen 2.5 Coder-3 B attains a recall of 0.68 on the vulnerable class --- substantially surpassing the 0.37 achieved by the LLaMA-3B fine-tuned model, thus exhibiting greater sensitivity to vulnerability cues at the expense of precision. Such a recall-oriented profile is desirable in scenarios where minimising false negatives outweighs concerns over false positives, for example during an initial triage phase in which flagged contracts are subsequently subjected to expert review.

\subsection{Comparative Analysis}
This section provides a comparative analysis of model-performance characteristics, delving into the resultant performance differences together with their practical deployment implications.

The improvements observed through fine-tuning, i.e. 19 percentage points for the LLaMA 3B fine-tuned model and 14 percentage points for the Qwen2.5Coder 3B fine-tuned model, demonstrate the effectiveness of domain-specific adaptation even with limited computational resources and training data. 

\begin{figure}[h]
\centering
\caption{Model Performance Improvement Through Fine-tuning}
\label{fig:performance_improvement}
\begin{tikzpicture}
\begin{axis}[
    ybar,
    bar width=0.6cm,
    width=\textwidth,
    height=6cm,
    legend style={at={(0.5,-0.2)}, anchor=north, legend columns=-1},
    ylabel={Accuracy (\%)},
    symbolic x coords={LLaMA 3B, Qwen2.5Coder 3B},
    xtick=data,
    nodes near coords,
    nodes near coords align={vertical},
    ymin=0, ymax=75,
    axis lines*=left,
]
\addplot coordinates {(LLaMA 3B, 48) (Qwen2.5Coder 3B, 45)};
\addplot coordinates {(LLaMA 3B, 67) (Qwen2.5Coder 3B, 59)};
\legend{Base Model, Fine-tuned Model}
\end{axis}
\end{tikzpicture}
\end{figure}
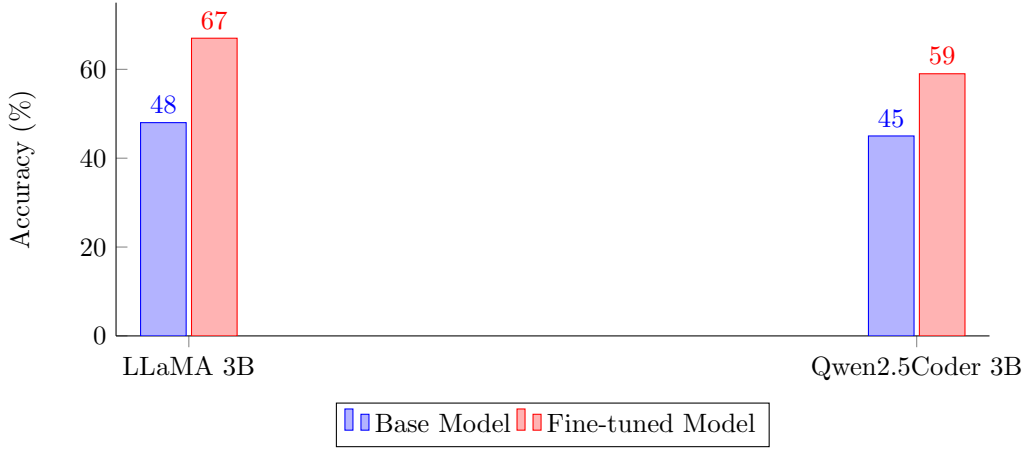

The larger performance gain realised by the fine-tuned LLaMA-3B model relative to the fine-tuned Qwen 2.5 Coder-3B model merits attention and likely stems from architectural factors that modulate fine-tuning effectiveness.  The magnitude of the observed performance uplift is striking in light of the task's inherent complexity and the constrained computational budget under which fine-tuning was conducted. 

The 8\% performance divergence between the fine-tuned LLaMA-3B (67\%) and Qwen 2.5 Coder 3B (59\%) models, necessitates a closer examination of architectural determinants. This gap underscores the pivotal influence of design choices and pretraining regimes on adaptation capacity for specialised niche tasks. Moreover, the fine-tuned LLaMA 3B model challenges the presumption that code-specialised models (such as the fine-tuned Qwen2.5Coder 3B model used)  are invariably optimal for all code-related applications inlcuding niche domains, particularly in resource-constrained settings.

Key architectural differences contribute to this performance differential:
\begin{itemize}
    \item \textbf{Positional Encoding:} LLaMA-3 B leverages Rotary Position Embeddings (RoPE) \cite{su2024roformer}, affording superior modelling of long-range dependencies—crucial for detecting cross-function reentrancy patterns—relative to the hybrid positional-encoding scheme adopted by Qwen 2.5 Coder.    
    
    \item \textbf{Pretraining Corpus:} LLaMA's more balanced pretraining corpus, with only 17\% code\footnote{\url{https://ar5iv.labs.arxiv.org/html/2407.21783}} versus Qwen2.5Coder's 70\%\footnote{\url{https://build.nvidia.com/qwen/qwen2_5-coder-7b-instruct/modelcard}}, which may contribute to the belief that diverse training data benefits transfer learning for reasoning-intensive tasks.
\end{itemize}

These findings suggest several implications:
\begin{itemize}
    \item \textbf{Domain-Specialized Models $\not\Rightarrow$ Task-Specialized Performance:} LLaMA's outperformance undermines the premise that code-specialised models are inherently superior for code and niche code domains.

    \item \textbf{Fine-tuning Efficiency:} The disparity in fine-tuning efficacy across architectures underscores that low-rank adaptation success is model-dependent.

    \item \textbf{Resource-Performance Tradeoffs:} The \% gap underscores the pivotal role of architectural selection under computational constraints, indicating that model choice can offset performance ceilings imposed by limited parameter budgets.
\end{itemize}

\subsection{Error Analysis}
A closer inspection of misclassification patterns furnishes salient insights into model behaviour and potential avenues for refinement. The fine-tuned LLaMA 3B model exhibits a pronounced propensity to classify contracts as non-vulnerable, attaining high specificity (92\%) yet limited sensitivity (37\%). In contrast, the fine-tuned Qwen 2.5 Coder 3B model delivers a more balanced error profile of 51\% specificity and 68\% sensitivity --- albeit with a lower overall accuracy.

This divergence in error patterns suggests that the models learned distinct feature representations during fine-tuning, specifically:
\begin{itemize}
    \item \textbf{Conservative Detection Heuristics:} The fine-tuned LLaMA 3B model appears to impose stricter detection thresholds.

    \item \textbf{Permissive Detection Criteria:} The fine-tuned Qwen2.5Coder 3B model demonstrated more lenient detection heuristics, improving sensitivity but increasing the false positive rate.
\end{itemize}

A contract-level examination of the misclassified instances reveals that both models struggled with the following cases:
\begin{enumerate}
    \item \textbf{Complex Cross-Contract Reentrancy Patterns:} Vulnerabilities spanning multiple contracts, typically manifesting through indirect state manipulation or multi-stage dependency chains.

    \item \textbf{Read-Only Reentrancy Patterns:} Subtle scenarios in which view functions introduce state inconsistencies that facilitate reentrancy.

    \item \textbf{Proxy and Delegatecall Implementations:} Contracts that employ sophisticated proxy patterns or invoke delegatecall, thereby introducing intricate control flows and obscuring state dependencies.
\end{enumerate}

These cases align with findings by Choi et al \cite{choi2021smartian}, who identified similar vulnerability patterns as particularly difficult for automated detection tools due to their intricate control flow and nuanced state management characteristics.

The fine-tuned LLaMA 3B model's abstention on 28 contracts embodies an emergent uncertainty --- instead of issuing low-confidence predictions, the model flags instances demanding expert scrutiny.

\subsection{Conclusions from the comparative evaluation of the fine-tuned LLaMA 3B and Qwen2.5Coder 3B models}

The comparative analysis of the fine-tuned LLaMA 3B and Qwen2.5Coder 3B models demonstrates several lessons for security and coding niche-oriented language-model applications:

\begin{enumerate}
    \item \textbf{Architectural Primacy for Security Tasks:} The performance disparity between architectures of equivalent parameter budgets indicates that architectural design exerts a substantial influence in niche tasks. This outcome may indicate that targeted architectural refinements may yield better performance relative to focusing on parameter expansion.

    \item \textbf{Complementary Error Characteristics:} The contrasting error profiles underscore the potential of ensemble strategies that harness complementary traits.

    \item \textbf{Uncertainty Recognition as a Feature:} The fine-tuned LLaMA 3B model's tendency to abstain from classifying contracts with ambiguous traits is advantageous in such critical settings, where over-confidence may conceal latent vulnerabilities.

    \item \textbf{Pre-training Diversity Trumps Domain Specialization:} The general purpose/not-fine-tuned models' performance indicates that a diverse pretraining corpus cultivates additional reasoning essential for vulnerability detection more effectively than a code-exclusive dataset --- which may be the case for other niche domains. This observation coincides with Kim et al's conclusion that transfer-learning efficacy on reasoning-intensive tasks hinges more on data diversity than on narrow domain specialisation \cite{kim2022modularized}.
\end{enumerate}

The observed performance profile indicates that parameter-efficient fine-tuning of model architectures furnishes a promising pathway to practical vulnerability-detection capabilities under constrained resource budgets which may also apply to other niche-domains.

\section{Comparing the Fine-tuned Small Models with a State-of-the-Art Larger Model}\label{sec:llmeval}

To put the work into perspective, we compared the fine-tuned small models with a state-of-the-art larger model, namely DeepSeek-r1 14B, possessing roughly 4.7 x the parameters of our fine-tuned LLaMA 3B model. The larger model delivered 70.43\% test accuracy --- only a 3.43\% gain over the fine-tuned LLaMA 3B model (67.39\%). The improvement, however, entailed markedly higher computational expenditure, and five smart contracts had to be omitted due to resource constraints. These findings resonate with Li et al's observation that properly tuned 1–3B parameter models may be able to attain 80–90\% of the task performance of models ten times their size \cite{li2023textbooks}. Table \ref{tab:comparison} shows the efficiency–performance trade-off, juxtaposing each model's accuracy, computational overhead, and contract-processing reliability.

\begin{table}[h!]
    \centering
    \caption{Model Performance and Computational Efficiency Comparison}
    \label{tab:comparison}
    \begin{tabular}{|l|c|c|c|}
        \hline
        \textbf{Model} & \textbf{Accuracy (\%)} & \textbf{Skipped Contracts} & \textbf{Parameters (B)} \\ \hline
        DeepSeek-r1 14B & 70.43 & 5 & 14 \\ \hline
        Fine-tuned LLaMA 3B & 67.39 & 28 & 3 \\ \hline
    \end{tabular}
\end{table}

\section{Conclusion}\label{sec:conc}
This paper provides several contributions to the smart contract analysis and small language-model adaptation areas:
\begin{itemize}
    \item \textbf{Small Model Adaptation:} We show that small language models (1–3B parameters) can be fine-tuned with parameter-efficient methods to detect reentrancy vulnerabilities in Solidity smart contracts, achieving substantial gains over baseline model performance --- 19 percentage points for the fine-tuned LLaMA 3B model and 14\% for the fine-tuned Qwen2.5Coder 3B model. This evidence affirms the practical viability of resource-constrained adaptation for specialised analysis.

    \item \textbf{Parameter-Efficient Fine-tuning:} We implement and assess a Low-Rank Adaptation (LoRA) fine-tuning regime, demonstrating that substantive task adaptation is attainable while updating fewer than 1\% of model parameters. This result extends to resource-bounded domain specialisation, evidencing that effective knowledge transfer can be realised without the computational overhead of full-parameter optimisation.

    \item \textbf{Synthetic Data Generation:} We devise and empirically validate a synthetic-data generation pipeline tailored to reentrancy vulnerabilities, thereby mitigating the acute data-scarcity constraint in vulnerability-focused AI-based techniques. 

    \item \textbf{Comparative Analysis of Model Architectures:} We conduct a comparison of divergent model architectures for niche code analysis, illuminating the interplay between pretraining objectives, architectural design, and specialisation capacity. The result that the fine-tuned LLaMA 3B model surpasses the code-specialised fine-tuned Qwen2.5Coder 3B warrants further investigation to challenge the assumption that domain-specific pretraining intrinsically yields superior performance for other niche domains.

\end{itemize}

Our results match that of Gunasekar et al \cite{li2023textbooks}, who show that properly fine-tuned small language models (1–3B parameters) can achieve 70–80\% of the performance of far larger models. This efficiency–performance trade-off marks an attractive point in the design space for practical niche applications.

\subsection{Future Research Directions}
We now lay out a number of future research directions:
\begin{itemize}
    \item \textbf{Scaling Model Size:} An immediate extension of this work is to interrogate the relationship between model scale and vulnerability-detection efficacy. Evaluating moderately larger architectures (7–13B parameters) under parameter-efficient fine-tuning could unlock additional gains while sustaining manageable resource demands, echoing Touvron et al's evidence that performance scaling persists in this parameter range for specialised tasks \cite{touvron2023llama}.

    \item \textbf{Ensemble Methods:} Combining outputs from multiple fine-tuned models constitutes a promising route to enhanced detection accuracy. Orthogonal error patterns from the fine-tuned LLaMA 3B and Qwen2.5Coder 3B models suggest complementary strengths that ensemble methods could harness. 

    \item \textbf{Uncertainty Quantification:} Introducing explicit uncertainty quantification would enhance the practical utility of language-model-driven vulnerability detection. Recasting the task as a ternary classification (e.g. vulnerable, non-vulnerable, or uncertain) or attaching calibrated confidence scores may better align model outputs with real-world assessment workflows.
    
\end{itemize}

\backmatter





\begin{appendices}





\end{appendices}


\bibliography{sn-bibliography}

\end{document}